\documentclass[]{ceurart}

\usepackage{listings}
\begin{document}

\copyrightyear{2025}
\copyrightclause{Copyright for this paper by its authors. Use permitted under Creative Commons License Attribution 4.0 International (CC BY 4.0).}

\conference{The 2nd Solid Symposium Poster Session, co-located with the 3rd Solid Symposium, April 24 -- 25, 2025, Leiden, Netherlands}

\title{Bridging the Discovery Gap in Solid Dataspaces with a Semantic Data Catalog}

\author[1]{Florian Hölken}[orcid=0009-0002-5191-901X, email=hoelken@uni-wuppertal.de]
\author[1]{Alexander Paulus}[orcid=0000-0002-9899-6736, email=paulus@uni-wuppertal.de]
\author[1]{Tobias Meisen}[orcid=0000-0002-1969-559X, email=meisen@uni-wuppertal.de]
\author[1]{André Pomp}[orcid=0000-0003-0111-1813, email=pomp@uni-wuppertal.de]

\address[1]{Institute for Technologies and Management of Digital Transformation, Lise-Meitner-Straße 27-31, 42117 Wuppertal, Germany}

\maketitle

\begin{abstract}
Dataspaces are an emerging concept to enable secure and controlled data sharing between multiple stakeholders. However, most existing dataspace architectures, such as those based on the concepts of IDSA (International Data Spaces Association) and Gaia-X, are heavily concept-driven and require significant effort to implement, making it difficult to quickly build a functional dataspace. Solid (Social Linked Data) offers a compelling alternative by providing a lighter and more practical approach. However, for Solid to function as a true dataspace, like traditional dataspaces, it must be built in a decentralized manner, i.e., the dataspace must consist of more than a single Solid server. However, this decentralization introduces a key challenge: Data Discovery. Unlike traditional dataspace architectures, which often include predefined components such as data catalogs to facilitate data discovery and access, Solid does not inherently provide such mechanisms. As a result, managing and locating relevant data across multiple Solid servers and pods becomes a significant barrier to usability. To address this gap, we present a FAIR-compliant data catalog for Solid. Our catalog follows the FAIR principles (Findability, Accessibility, Interoperability, and Reusability) by integrating semantic models that describe all available data sources within the Solid ecosystem. This approach enables enhanced data discovery, improves accessibility through standardized metadata, enhances interoperability between different data sources, and promotes data reuse across applications. By bridging this critical gap, our catalog improves the usability of Solid-based dataspaces, making them a viable option for decentralized and open data ecosystems, such as smart city applications.
\end{abstract}
\section{Introduction}
The concept of dataspaces has emerged as a response to the increasing need for decentralized, interoperable, and secure data-sharing ecosystems. Unlike traditional centralized data platforms, dataspaces provide a framework where multiple independent stakeholders can share and access data under defined governance models while maintaining control over their respective data sources \cite{Curry2022}. This approach is particularly relevant in domains such as smart cities, where heterogeneous data from various public and private entities must be efficiently shared and utilized.

However, most existing dataspace architectures, such as IDSA and Gaia-X, are heavily conceptual and complex. They emphasize trust frameworks, policy enforcement, and governance mechanisms, but their technical implementations often require significant effort and infrastructure. Setting up a functional dataspace within these ecosystems is not trivial — it involves establishing a network of trusted entities, integrating identity and access management systems, and deploying dedicated service components like metadata registries and data marketplaces \cite{Solmaz2022}. As a result, these architectures are primarily suited for enterprise-level data ecosystems, where formalized governance structures and extensive resources are available.

In contrast, Solid provides an alternative approach to decentralized data sharing. Originally conceived by Tim Berners-Lee, Solid focuses on personal data sovereignty, allowing individuals and organizations to store their data in personal data pods while enabling controlled access through standard web technologies. By using Linked Data principles, Solid ensures interoperability across different data sources without requiring a central authority. This makes it an attractive candidate for building lightweight and rapidly deployable dataspaces, particularly in domains where smaller, more flexible data-sharing ecosystems are needed \cite{Sambra2016}.

While Solid offers a promising foundation for decentralized dataspaces, it introduces a critical challenge: \textit{data discovery}. Unlike traditional dataspace architectures, which include predefined components such as data catalogs and metadata registries to enable structured data discovery, Solid does not natively provide such mechanisms. This is especially problematic in a fully decentralized Solid ecosystem, where multiple independent Solid servers host fragmented data across different participants as needed for dataspaces. Without a way to efficiently index, describe, and discover available datasets, the practical usability of a Solid-based dataspace remains limited.

In conventional dataspace implementations, data findability is addressed through centralized or federated catalog services, which maintain metadata descriptions of all available datasets within the ecosystem \cite{Jahnke2024}. These catalogs enable semantic search, data governance, and automated interoperability by leveraging standardized metadata schemas and ontologies. Without an equivalent component, Solid dataspaces risk becoming fragmented, making it difficult for users and applications to locate relevant datasets efficiently.

To bridge this gap, we propose a FAIR-compliant semantic data catalog for Solid-based dataspaces. Existing dataspace architectures like IDSA or Gaia-X rely on centralized or federated infrastructures with components such as identity providers, service registries, and metadata catalogs, which are designed for enterprise-level ecosystems with formal governance structures. In contrast, Solid is inherently decentralized and focused on individual data ownership, lacking built-in mechanisms for federated data discovery and metadata management across multiple pods. As a result, directly applying traditional dataspace components to Solid is impractical, creating the need for a lightweight, decentralized solution tailored to Solid’s unique architecture. By incorporating semantic models and metadata standards, our catalog enables:
\begin{itemize}
    \item \textbf{Findability:} Indexing and organizing available data sources across multiple Solid servers.
    \item \textbf{Accessibility:} Linking datasets with their respective Solid pods and corresponding data owner.
    \item \textbf{Interoperability:} Standardized metadata schemas to ensure seamless data integration.
    \item \textbf{Reusability:} Semantic models as semantic annotations that enhance data contextualization and improve usage.
\end{itemize}

Our approach aims to enhance the practicality of Solid-based dataspaces by providing a structured and scalable solution for data discovery. By integrating a semantic data catalog, we enable more efficient data sharing and governance, making Solid a viable option not just for individual data sovereignty, but for broader, multi-stakeholder data ecosystems such as those that are needed for smart cities or open data initiatives.

\section{Related Work}
Data catalogs have emerged as essential tools for metadata management, supporting data discovery, integration, and governance. In general, these systems curate and maintain metadata, creating structured inventories of data assets \cite{Ehrlinger2021}. Centralizing metadata bridges the gap between data producers and consumers, enhancing data usability \cite{Subramaniam2023,Shanmugam2016}. Various classifications of data catalogs have been proposed based on application types and integration capabilities. Jahnke and Otto \cite{Jahnke2023} distinguish between internal and cross-enterprise catalogs, highlighting the need for tailored solutions. Beyond traditional metadata management, semantic technologies have been integrated into data catalogs. The Semantic Data Dictionary (SDD) approach formalizes metadata representation by annotating datasets with ontology-based concepts. This method transforms raw tabular data into knowledge graphs, enhancing data discoverability, traceability, and integration, in line with FAIR principles \cite{Rashid2020}. Compared to conventional data dictionaries, SDDs improve metadata standardization and machine-readability. Semantic layers have also been utilized in large-scale data infrastructures, such as data lakes. Dibowski et al. \cite{Dibowski2020} introduce an ontology-based semantic layer within the Bosch Automotive Data Lake. Their approach incorporates a Data Catalog Provenance and Access Control (DPAC) ontology, enabling semantic search, provenance tracking, and granular access control. These capabilities demonstrate the effectiveness of semantic technologies in managing complex, heterogeneous data environments.

\section{The Semantic Data Catalog}
To ensure the continuous growth and adaptability of the semantic data catalog within a Solid-based dataspace, a structured approach is required for integrating new datasets. The onboarding process consists of three key steps: registration of Solid pods, datasets, and metadata enrichment of datasets using semantic modeling, which collectively enable seamless data discovery and interoperability.

\subsection{Adding Solid pods}
When a new Solid data pod is added to the dataspace, it must first register with the semantic data catalog. This process is supported by Solid's WebID verification, which authenticates the user by logging them into their pod, enabling them to add datasets to the catalog while ensuring trust and governance. For our implementation and testing, we utilized the Solid Community Server (SCS) as a reference environment for hosting and managing Solid pods. In our tests, we deployed three Solid Community Servers, hosting a total of ten individual pods across these servers.

\subsection{Adding Datasets}
Once a Solid pod has been registered, datasets from the corresponding pod can be added to the semantic data catalog. This is done through the catalog interface, where the user provides metadata and selects datasets from their pod. The user also assigns the corresponding semantic model.

A key challenge in data indexing is the lack of standardization in metadata representation. To address this, we employ the DCAT (Data Catalog Vocabulary) standard developed by W3C. DCAT provides a structured approach to metadata representation, supporting data discovery and alignment with dataspace guidelines. It is widely adopted for cataloging datasets, distributions, and services within data ecosystems \cite{w3c2024dcatv3}. While DCAT forms the backbone of the metadata structure, it is extended with semantic models to capture contextual information about data assets.

\begin{figure}[t]
    \centering
    \includegraphics[width=\textwidth]{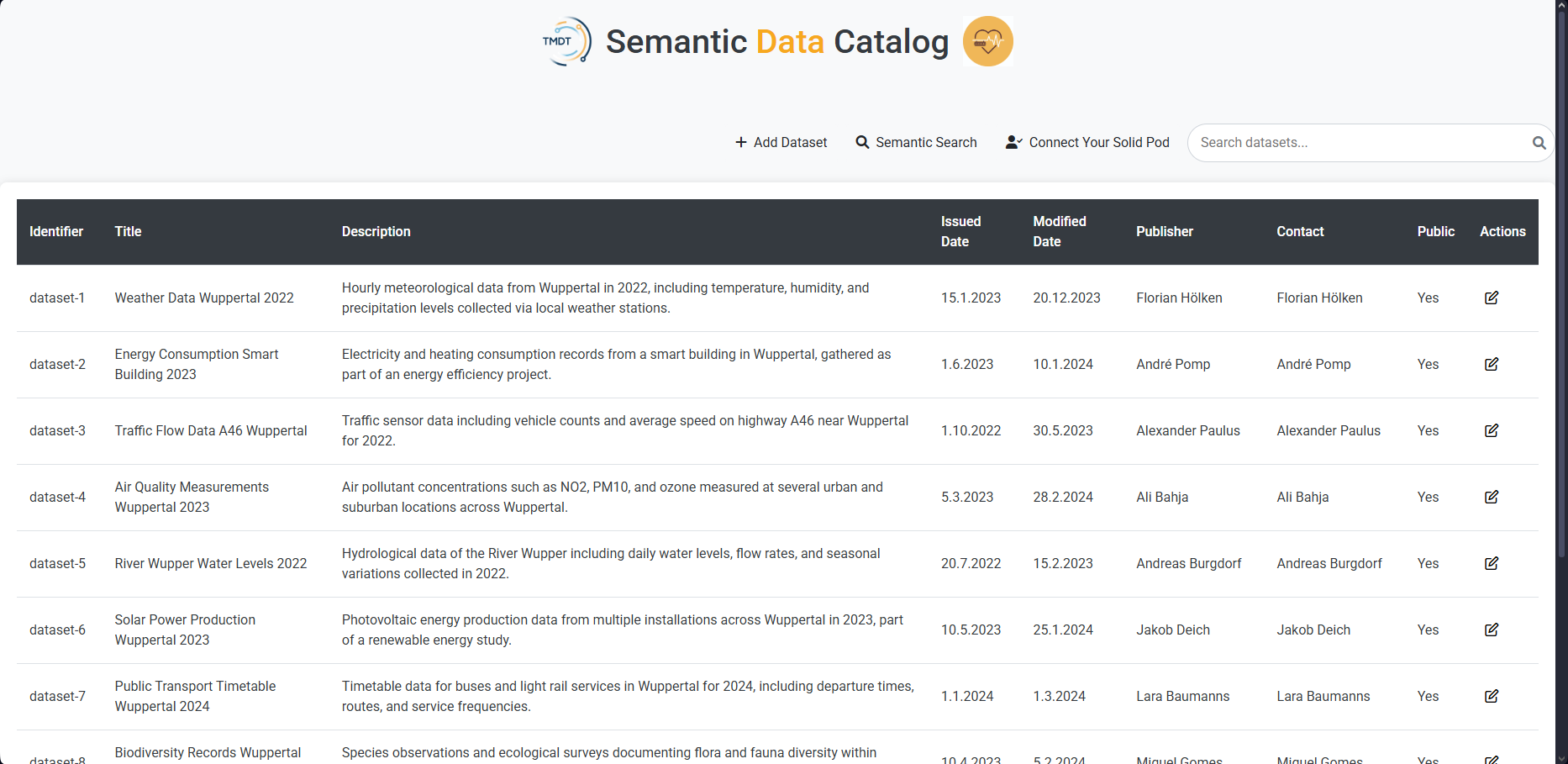}
    \caption{Screenshot of the semantic data catalog showing different indexed datasets and their metadata.}
    \label{fig:semantic_data_catalog}
\end{figure}

To implement our catalog solution in compliance with DCAT, we initially structured the data assets around datasets and distributions as the primary organizational units. An RDF (Resource Description Framework) description of the catalog and all registered datasets is provided through the web interface, with metadata fully expressed using the appropriate DCAT classes and properties, following the semantics defined by the specification. While this ensures baseline compliance with DCAT, a current limitation of our implementation is the absence of additional DCAT components such as data services and dataset series, which will be integrated in future development stages to further enhance interoperability within the dataspace. The catalog interface provides a user-friendly view where registered datasets, along with their enriched metadata, are visualized and managed (see Figure~\ref{fig:semantic_data_catalog}).

\subsection{Semantic Modeling}
To ensure interoperability across heterogeneous datasets, we integrate semantic models, as outlined by Knoblock et al. \cite{Taheriyan2016,Knoblock2019}. These models, expressed in RDF, enrich metadata with additional layers of meaning, allowing users to understand relationships, provenance, and constraints of the datasets. This semantic enrichment ensures a coherent and queryable data structure while maintaining flexibility in dataset representation.

\begin{figure}[t]
    \centering
    \includegraphics[width=\textwidth]{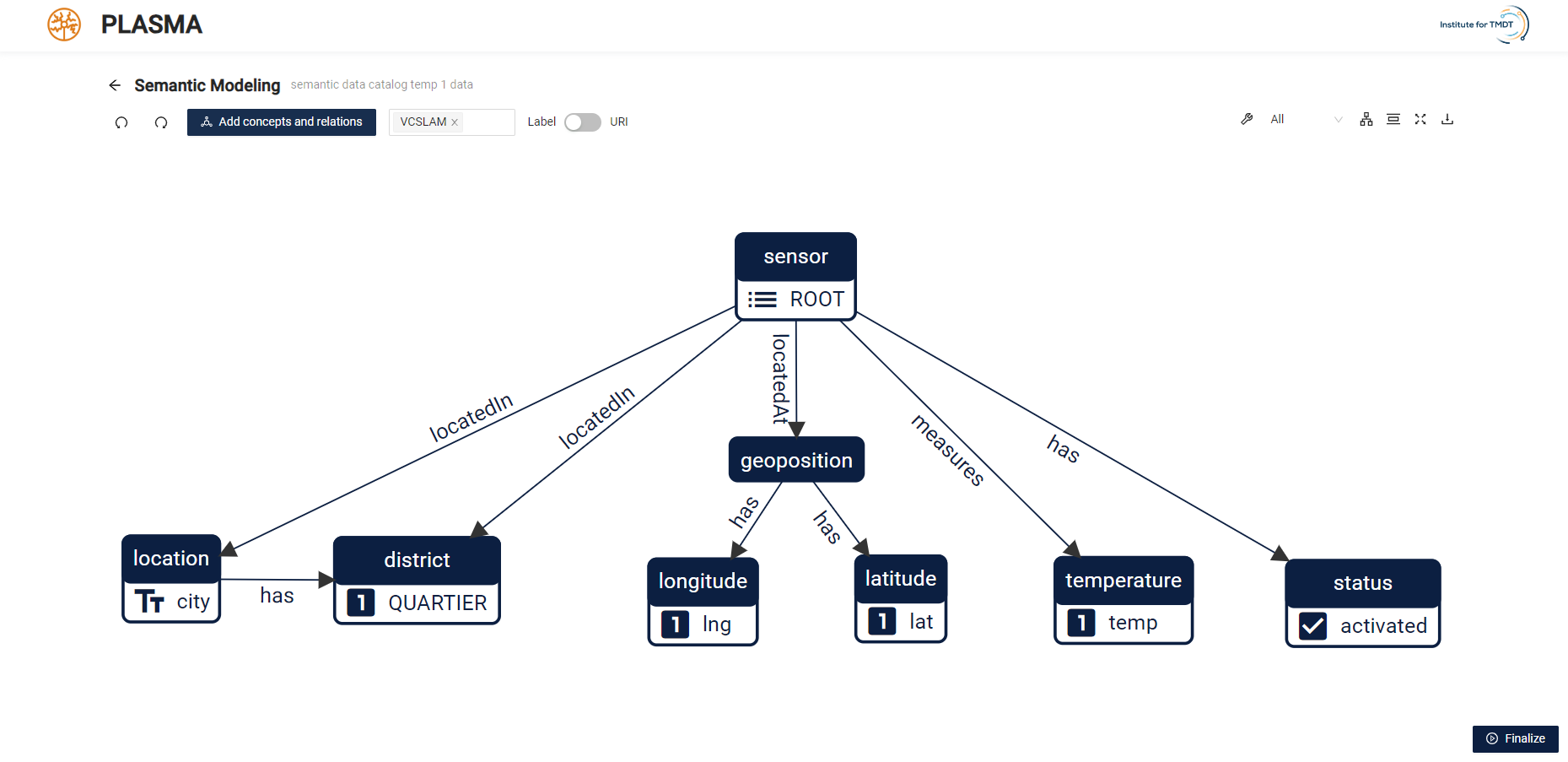}
    \caption{Screenshot of a semantic model created with PLASMA.}
    \label{fig:semantic_model}
\end{figure}

For the semantic modeling process, we employ PLASMA \cite{Paulus2023}, a tool specifically designed to assist in the creation of semantic models. PLASMA combines automated and semi-automated techniques for generating semantic models and offers a user-friendly interface for mapping dataset attributes to ontologies (see Figure~\ref{fig:semantic_model}).

\section{Conclusion and Future Work}
In this work, we presented a FAIR-compliant semantic data catalog to address the data discovery challenge in decentralized Solid-based dataspaces. By integrating DCAT as the metadata standard and enriching datasets with semantic models, our solution enhances findability, interoperability, and reuse of data assets. However, the current implementation has several limitations. At present, the catalog only supports public datasets, limiting its applicability in scenarios requiring controlled access to private data. Additionally, the catalog focuses solely on datasets and distributions as organizational units, with other DCAT components, such as data services and dataset series, yet to be implemented. Future work will address these gaps by introducing mechanisms to handle private datasets through request management and access control, as well as extending DCAT compliance to include additional resource types. We also plan to integrate a SPARQL endpoint to leverage the full potential of DCAT metadata and semantic models, enabling semantic search capabilities and overall data discovery within the dataspace. Moreover, the inclusion of semantic models enables future research into semantic processing, such as schema alignment and ontology-based data integration, to further enhance interoperability between heterogeneous data sources.

\section*{Declaration on Generative AI}
During the preparation of this work, the author(s) used GPT-4o and DeepL Write for grammar and spelling checks. After using these tool(s)/service(s), the author(s) reviewed and edited the content as needed and take(s) full responsibility for the publication’s content.

\section*{Acknowledgements}
This work has been supported as part of the research project \emph{Gesundes Tal} in collaboration with the city of Wuppertal, funded by the Federal Ministry of Housing, Urban Development and Building (BMWSB) and the Reconstruction Loan Corporation (KfW) through the funding program “Modellprojekte Smart Cities: Stadtentwicklung und Digitalisierung” (grant number 19454890).

\bibliography{bibliography}

\end{document}